\documentclass[useAMS,usenatbib]{mn2e}

\usepackage{graphicx}
\usepackage[fleqn]{amsmath}

\newcommand{\mpl}{M_{\mathrm p}}

\newcommand{\rp}{r_{\mathrm p}}
\newcommand{\pp}{\varphi_{\mathrm p}}
\newcommand{\op}{\Omega_\mathrm{p}}
\newcommand{\okep}{\Omega_\mathrm{K}}
\newcommand{\cs}{c_\mathrm{s}}
\newcommand{\me}{\mathrm{M_\oplus}}
\newcommand{\xs}{x_\mathrm{s}}

\newcommand{\be}{ \begin {equation}}
\newcommand{\ee}{ \end {equation}}

\title[On the corotation region for low-mass planets]{On the width and shape of the corotation region for low-mass planets}
\author[S.-J. Paardekooper and J.C.B. Papaloizou]{S.-J. Paardekooper$$\thanks{E-mail:
S.Paardekooper@damtp.cam.ac.uk} and J.C.B. Papaloizou\\
DAMTP, University of Cambridge, Wilberforce Road, Cambridge CB3 0WA, United Kingdom}
\begin{document}

\date{Draft version \today}

\pagerange{\pageref{firstpage}--\pageref{lastpage}} \pubyear{2008}

\maketitle

\label{firstpage}

\begin{abstract}
We study the coorbital flow for embedded, low mass planets. We provide
 a simple semi-analytic model for the corotation region, which is
 subsequently compared to high resolution numerical simulations. The
 model is used to derive an expression for the half-width of the
 horseshoe region, $\xs,$ which in the limit of zero softening is
 given by  $ \xs/\rp = 1.68(q/h)^{1/2}$, where $q$ is the planet to
 central star mass ratio, $h$ is the disc aspect ratio and $\rp$ the
 orbital radius. This is in very good agreement with the same quantity
 measured from  simulations. This result is used to show that
 horseshoe drag is about an order of magnitude larger than the linear
 corotation torque in the zero softening limit. Thus the horseshoe
 drag, the sign of which depends on the gradient of specific
 vorticity, is  important for estimates of the  total torque acting on
 the planet. We  further show that   phenomena, such as the Lindblad
 wakes,  with a radial separation   from corotation of $\sim$ a
 pressure scale height  $H$ can affect $\xs$, even though for low-mass
 planets $\xs \ll H$. The effect is to distort streamlines and to
 reduce $\xs$ through the action of  a back pressure. This effect is
 reduced for smaller gravitational softening parameters and planets of
 higher mass, for which $\xs$ becomes comparable to $H$. 
\end{abstract}



\section{Introduction}
Immediately after the discovery of the first extrasolar planet
\citep{1995Natur.378..355M}, a Jupiter-mass planet in a very close
orbit, it was realised that this class of planets, the Hot Jupiters,
could not have been formed at their present location. In stead, they
should have formed further out in the protoplanetary disc, and
migrated inward afterwards. As outlined in
\cite{1979ApJ...233..857G,1980ApJ...241..425G}, planets embedded in
protoplanetary discs indeed will undergo orbital evolution through
disc tides, and a great deal of theoretical work has been dedicated to
understand the direction and magnitude of planetary migration
\citep[for an overview see][]{2007prpl.conf..655P}. 

One can distinguish three types of migration. High mass planets,
comparable to Jupiter, open up deep gaps in their discs, after which
they migrate on approximately a viscous time scale
\citep{1986ApJ...309..846L}. This is called Type II migration
\citep{1997Icar..126..261W}. Less massive planets, comparable to
Saturn, embedded in massive discs may undergo fast Type III migration
\citep{2003ApJ...588..494M,2008MNRAS.386..179P}. Both Type II and Type
III migration may be directed inward or outward, depending on local
conditions in the disc
\citep{2007MNRAS.377.1324C,2008MNRAS.387.1063P}, but the general trend
is inward migration. 

Low-mass planets, up to a few times the mass of the Earth ($\me$),
 undergo Type I migration \citep{1997Icar..126..261W}. 
This type of migration is driven by a linear wave response in the
 disc, leading to a characteristic two-armed spiral pattern
 \citep{2002MNRAS.330..950O}. The waves can be understood to be
 excited at Lindblad resonances \citep{1979ApJ...233..857G} and lead
 to a torque on the planet that is due to asymmetries in the density,
 pressure and rotation profile in the disc
 \citep{1997Icar..126..261W}. The resulting migration direction is
 inward for all reasonable disc parameters
 \citep{1993Icar..102..150K,2002ApJ...565.1257T}.  

Apart from this wave, or Lindblad, torque, embedded planets are also
subject to corotation torques \citep{1979ApJ...233..857G}. Two
descriptions of the corotation torque exist in the literature. As
advocated by \cite{1979ApJ...233..857G}, one can perform a linear
analysis of the corotation resonance, which leads to a torque
proportional to the radial gradient of specific vorticity, or
vortensity, in the unperturbed disc. Semi-analytical and numerical
studies lead to expressions for the total corotation torque
\citep{1993Icar..102..150K,2002ApJ...565.1257T} in two- and in
three-dimensional discs. We will refer to this torque as the linear
corotation torque.  

A different view on the corotation torque was given by
\cite{1991LPI....22.1463W}, who considered the torque due to material
near the orbit of the planet that executes horseshoe turns. The total
corotation torque is again proportional to the radial gradient of
specific vorticity in the unperturbed disc, but the model contains a
free parameter $\xs$, the width of the horseshoe region. We will refer
to this torque as the horseshoe drag. The relation between the two
descriptions has never been clarified so far. 

Linear theory has been compared successfully against numerical
hydrodynamical simulations, in 2D
\citep{2002A&A...385..647D,2004MNRAS.350...849N} as well as in 3D
\citep{2003ApJ...586..540D}. At this point we remark, however, that in
these studies, only discs with small gradients in specific vorticity
were considered, that is, cases where the corotation is supposed to be
weak.  For shallow surface density gradients, resulting in strong
corotation torques, intermediate mass planets may reverse their
direction of migration \citep{2006ApJ...652..730M}. 

All studies mentioned so far have made use of the simplifying
assumption that a barotropic (or isothermal) equation of state
applies, in which case no energy equation needs to be solved. The
dramatic effects of releasing this assumption were first noted in
\cite{2006A&A...459L..17P}, where it was shown that low-mass planets
can migrate \emph{outward} in non-isothermal discs. Subsequently it
was realised that this was due to a radial entropy gradient in the
unperturbed disc \citep{2008A&A...478..245P}, giving rise to a strong,
positive corotation torque. \cite{2008ApJ...672.1054B} provided a
linear analysis of the problem, and argued that a linear effect due to
a background entropy gradient can be strong enough to reverse the
torque on low-mass planets. However, \cite{2008A&A...485..877P} showed
that this linear effect is in fact small, and that a non-linear effect
is responsible for the torque reversal as seen in
\cite{2006A&A...459L..17P}.   

The non-linear torque, as studied in \cite{2008A&A...485..877P}, is
 closely related to the idea of horseshoe drag originally introduced
 by by \cite{1991LPI....22.1463W} as both of these are produced by
 disc material undergoing horseshoe turns in the neighbourhood of the
 planet. For this reason it is  very important to have a clear
 understanding of the horseshoe drag and in particular its
 relationship to linear corotation torques that are often used to
 estimate torques arising from coorbital effects. In this paper, we
 will provide an analysis of the horseshoe region for low-mass
 planets, in particular we determine  its half-width $\xs$ for
 softening lengths ranging between zero and the disc scale height. For
 simplicity we shall return to considering  a barotropic equation of
 state and work with a two dimensional model. Since the horseshoe drag
 is proportional to $\xs^4$ \citep[see][]{1991LPI....22.1463W}, a good
 estimate of this parameter is critical in determining the total
 torque on the planet which is found in  to be significantly larger
 than estimates based on linear theory for zero softening. In an
 accompanying paper, we perform a general  study on the behaviour of
 the torques and their dependence on other parameters such as the disc
 viscosity.  

The plan of this paper is as follows. In Section \ref{secEq}, we
 review the basic equations and introduce our local model. In Section
 \ref{secCor}, we study the structure of the corotation region, and in
 Section \ref{secStream} we analyse the resulting streamlines. These
 are then compared with numerical hydrodynamic simulations in Section
 \ref{secNum}, after which we give a brief discussion and conclusions
 in Section \ref{secDisc}.  


\section{Basic equations and disc models}
\label{secEq}
The basic equations are those of the conservation of mass, momentum
and energy for a two dimensional  disc in a  frame rotating with
angular velocity $\op.$ we adopt a cylindrical polar coordinate system
$(r,\varphi, z )$ with origin $(r=0)$  located at the central
mass. The disc then occupies the plane $z=0.$ The continuity equation
and the equation of motion take  the form 
\be \frac{\partial \Sigma}{ \partial t}= -\nabla\cdot(\Sigma  
{\bf v})\label{cont}
\ee
and 
\be \frac{D {\bf v}}{D t} +2\op{\hat {\bf k}} \times {\bf v}
=-\frac{1}{\Sigma }\nabla \Pi - \nabla\Phi \label {mot}\ee
respectively. Here, $\Sigma $ denotes the surface density, ${\bf v} =
(v_r, v_{\varphi})$ the velocity, ${\hat {\bf k}}$ denotes the unit
vector in the vertical direction, and $\Pi $ the vertically integrated
pressure. Thus
\be \Pi = \int^{\infty}_{-\infty}Pdz.\ee
The total potential $\Phi$  is taken to be $\Phi =\Phi_\mathrm{G}
+\op^2r^2/2,$ where $\Phi_\mathrm{G}$ is  the gravitational
potential. The convective derivative  is defined by  
\be \frac{D}{Dt} \equiv  \frac{\partial}{\partial t}+ {\bf v}\cdot \nabla.\ee
We adopt a  barotropic equation of state such that
\begin{equation}
\Pi =F(\Sigma ),
\end{equation}
with $F(\Sigma )$ being a prescribed function of $\Sigma .$ The square
of the sound speed is given by 
\begin{equation}
 \cs^2 =\frac{dF(\Sigma)}{d\Sigma}.
\end{equation}
When a  power law is adopted  such that 
 \be F = K\Sigma^{\beta}, \label{EOS}\ee
with $K$ and   $\beta$  being   constants, $\cs=\beta (\Pi/\Sigma).$
 The discs we consider are assumed to be low mass with the Toomre
 parameter $Q= (\Omega \cs)/(\pi G \Sigma) \gg 1.$ The self-gravity of
 the disc is accordingly  neglected. 

\subsection{Global studies}
The gravitational potential is assumed to be due to the central mass
and perturbing planet such that $\Phi_\mathrm{G} = \Phi_\mathrm{G0}+
\Phi_\mathrm{Gp},$  when the $z$ dependence is neglected. These are
given by 
\be \Phi_\mathrm{G0} = \frac{-GM_*}{r}, \ee
and
\begin{eqnarray}
\Phi_\mathrm{Gp} = 
\frac{-G\mpl}{\sqrt{r^2+\rp^2-2r\rp\cos(\Delta\varphi)+b^2\rp^2}}+\nonumber\\  
\frac{G\mpl r \cos(\Delta \varphi)}{\rp^2},\label{pot}
\end{eqnarray}
where  $\Delta \varphi = \varphi -\pp.$  In the above $M_*$ denotes
 the mass of the central object, with  $\mpl,$ $\rp,$ and  $\pp$
 denoting  the mass, orbital radius and angular coordinate  of the
 protoplanet respectively. The  gravitational softening parameter  is
 $b.$ The last term in (\ref{pot}) is the indirect term which accounts
 for the gravitational acceleration of the origin of coordinate system
 due to the action of the protoplanet. 

Rather than neglecting the vertical dependence of the perturbing
 potential, one may adopt a vertical average based on the density,
 $\rho,$ thus one replaces the expression (\ref{pot}) by its vertical
 average 
 \be \Phi_\mathrm{Gp} \rightarrow {1\over \Sigma}\
\int^{\infty}_{-\infty}\rho \Phi_\mathrm{Gp}dz. \label{pot1}\ee
This procedure is more consistent with the view that the two
 dimensional disc representation should be derived from applying a
 vertical averaging procedure and accordingly provides a vertical
 average. On the other hand if vertical motions near the midplane are
 small, use of the unaveraged form may be appropriate for representing
 the structure there as in  a stacked layer model
 \citep[eg.][]{2006ApJ...652..730M}. 

When $\op=0$ the reference frame is non rotating but non inertial as
the origin accelerates together with the central mass due to the
action of the protoplanet. Numerical calculations are most
conveniently performed in a frame corotating with the
protoplanet. Then $\op$ becomes the circular Keplerian angular
velocity at radius $\rp.$ At a general radius, $r,$ the Keplerian
angular velocity is given by $\okep(r) = (GM_*/r^3)^{1/2}.$ Thus
$\okep(\rp) = \op.$ The local vertical scale height of the disc is
then defined  through $H = \cs/\okep.$ Hydrostatic equilibrium in the
$z$ direction implies that 
\be \rho = \rho_0\exp{\left[-\int_0^{z}(z'/H^2)dz'\right]},\ee 
where $\rho_0$ is the midplane density. The value of $H$ at the disc
midplane is a measure of the local  disc semi-thickness. From now on
$H$ will be  stand for this quantity. 

\subsection {Local Description}
As we are interested in a local region close to the planet, it is
useful to work with a form of the basic equations adapted for this
purpose. We use the well known shearing box formalism
\citep{1991LPI....22.1463W}. This uses a Cartesian coordinate system
$(x, y, z)$  corotating with the protoplanet and with origin at the
centre of the planet. The disc velocity in the unperturbed state with
no protoplanet, is  in the local approximation,${\bf v}= (v_x,v_y,0)=
{\bf v}_0 = (0, -3\op x/2,0),$ corresponding to a linear shear. When a
protoplanet is present the equation of motion may be written 
\be {\partial{\bf v} \over \partial t} + {\bf v}\cdot\nabla {\bf v} +
2\op{\hat {\bf k}} \times {\bf v}
=-{1\over \Sigma }\nabla \Pi - \nabla\Phi_\mathrm{L} \label {motloc},\ee
where $\Phi_\mathrm{L} = \Phi_\mathrm{Gp} - 3\op^2x^2/2.$ The
protoplanet potential is here taken to be given by 
\be \Phi_\mathrm{Gp} = {-G\mpl\over \sqrt{x^2 + y^2 +b^2r_p^2}}. \ee
The indirect term being small compared to the direct potential close
to the planet is neglected. The continuity equation remains of the
same form as equation (\ref{cont}).

\section{The corotation region}
\label{secCor}
We now consider the structure of the corotation region. We begin by
considering steady state solutions of the basic equations. In this
case the two components of the equation (\ref{motloc}) are 
\be  {\bf v}\cdot\nabla v_x -2\op v_y
=  - {\partial(w+\Phi_\mathrm{L})\over \partial x} \label {motlocx},\ee
and
\be  {\bf v}\cdot\nabla v_y + 2\op v_x
=  - {\partial(w+\Phi_\mathrm{L})\over \partial y} \label {motlocy} ,\ee
where we have introduced the enthalpy, $w(\Sigma),$ which is defined
through $dw/d\Sigma = \cs^2/\Sigma$ together with the specification
that $w(0) =0.$ 

\subsection{A simple one dimensional model}
We now simplify the problem by in the first instance neglecting the
degree of freedom corresponding to epicyclic motions. These are not
expected to play a major role in the horseshoe region. One expects a
balance between the Coriolis force and potential gradient with ${\bf
  v}\cdot\nabla v_x$ being negligible in equation
(\ref{motlocx}). Neglecting this term is precisely the approximation
often used in celestial mechanics to enable the derivation of a second
order differential equation  describing  particle motion on horseshoe
orbits. It breaks down only close to the protoplanet. In the next
subsection, we discuss a more complete description of the coorbital
region.  

Adopting this approximation and using  equation (\ref{motlocx}) we set 
\be v_y =  {1\over 2 \op} {\partial \chi \over \partial x}
= -{1\over  \Sigma } {\partial \psi \over \partial x}, \label{chidef}\ee
where  $\chi = w+\Phi_\mathrm{L},$  with $\psi$ being the stream
function. Equations (\ref{motlocx}) and (\ref{motlocy}) may also be
written in the form 
\be 
\left (2\op + {\partial v_y \over \partial x}\right){\hat {\bf k}}
\times {\bf v} = - \nabla\left({1\over 2}v_y^2 + w
+\Phi_\mathrm{L}\right) \label {vortcon}.\ee 
From this and the steady state form of the continuity equation
(\ref{cont}) it follows that both 
$$E_0(\psi) = \left({1\over  2}
v_y^2 + w +\Phi_\mathrm{L}\right) \hspace{0.5cm} {\rm and}
\hspace{0.5cm} \xi_0(\psi) = {\left(2\op + {\partial v_y \over
\partial x }\right)\over \Sigma} $$ 
are constant on streamlines. These are statements of the conservation
of the Bernoulli constant and the specific vorticity or vortensity on
streamlines under the approximation used here which has the
consequence that the contribution from the radial velocity is
neglected.  We  note that the functional forms of $E_0(\psi)$ and
$\xi_0(\psi)$ cannot be determined from the inviscid equations but
have to be prescribed externally. From (\ref{chidef}) it then follows
that  
\be \frac{1}{2\op}\frac{\partial^2 \chi}{\partial x^2}+2\op=
\xi_0\Sigma .\ee
Using the fact that $w(\Sigma) = \chi -\Phi_\mathrm{L}$ to eliminate
$\Sigma,$ a single equation for $\chi$ then results. We note that when
a power law equation of state is used with $\beta =2,$ see
(\ref{EOS}), $w(\Sigma) = 2K\Sigma,$ 
and we have the very simple relation 
\be c^2_s = \chi -\Phi_\mathrm{L}.\label{ENTH}\ee
\noindent We thus obtain  an  equation for
$\chi$ of the form 
\be \frac{1}{2\op}\frac{\partial^2 \chi}{\partial x^2}+2\op=
\frac{\xi_0\Sigma}{\cs^2}( \chi -\Phi_\mathrm{L}) .\ee
Setting $\chi = Y -3\op^2 x^2/2,$ this leads to
\be  Y =\Phi_\mathrm{Gp} +
\frac{\cs^2}{2\xi_0\Sigma\op}\frac{\partial^2 Y}{\partial x^2} +
\frac{\cs^2\op}{2\xi_0\Sigma}. \label{cogov}\ee 
We comment that although we adopted a power law equation of state with
$\beta =2$  to obtain (\ref{cogov}), because $w$ is a linear function
of $\Sigma$ in this case, the same equation applies to any barotropic
equation of state in the regime where the surface density variations
are small enough that only first order variations in the equation of
state need to be considered.  Then $\cs^2/\Sigma$ is taken to be  the
constant background value. Of course when $\beta =2,$ $\cs^2/\Sigma =
2K = \hspace{0.2cm} {\rm constant}$ regardless. 

\subsubsection{Solution for the coorbital region}
Equation (\ref{cogov}) contains no $y$ derivatives and can accordingly
be solved as a second order differential equation. Adopting the
boundary condition that $Y $  is bounded for $|x|\rightarrow \infty$
the solution can be written down in the standard form 
\be Y =-\int^{\infty}_{-\infty}
G(x,x')\left(\op^2 +\Phi_\mathrm{Gp}(x',y) 
\left({2\xi_0\Sigma\op \over \cs^2}\right)\right)dx' \label{GRN1D},\ee
where  unless explicitly stated, quantities in this and other
integrands are evaluated at $(x',y')$ and the Green's function
satisfies 
\be \frac{\partial^2 G(x,x')}{\partial x'^2} -
G(x,x')\left(\frac{2\xi_0\Sigma\op}{\cs^2}\right)_{x=x'}=
 \delta(x-x') . \label{GRNEQ1}\ee
with $G(x,x')\rightarrow 0$ for $|x'|\rightarrow \infty.$ Having found
$Y=\chi +3\op^2x^2/2,$ the streamlines can be found from 
\be
E_0(\psi) = {1\over  8\op^2} \left(\frac{\partial \chi}{\partial
  x}\right)^2 + \chi \label{STRML}\ee 
which is constant on them. 

\subsubsection{Constant vortensity}
The corotation region is generally of small radial extent so that the
quantity  $\xi_0,$ representing the variation of vortensity should be
approximately constant unless the profile is very sharp. Therefore for
reasonable smoothly varying cases vortensity variation is not expected
to have much  effect on the the coorbital structure except possibly
for large softening cases perturbed by the density wakes associated
with Lindblad torques (see below). Accordingly we shall specialise to
the case of constant vortensity while retaining  $\beta=2,$ so that
both $\Sigma / \cs^2$ and  $\xi_0$  are constant. By considering large
distances from the protoplanet, we infer that $\xi_0
=\op/(2\Sigma_0),$ where $\Sigma_0$ is the uniform surface density at
large distances from the protoplanet. Determination of the Green's
function is straightforward in this case. We obtain 
\be  G(x,x') = - {1\over 2|k|}\exp{(-|k||x-x'|)},\ee
where  
\be k^2 = {\op^2 \over c_\mathrm{s0}^2} \equiv H^{-2},\ee
where $c_\mathrm{s0}$ is the uniform sound speed at large distances
from the protoplanet and $H$  is the scale height.

Although solution of (\ref{cogov}) is straightforward, we comment that
the scale of the decay of the Green's function is the scale height and
thus phenomena that distance away from the protoplanet and not
included here,  such as the prominent wakes associated with Lindblad
torques, can distort the flow (see below). In principle this effect
could be included through the boundary conditions on $G$ but we shall
not consider it further in this section. 

\subsubsection{Asymptotic series and softened gravity model}
\label{asymp}
We note also that in this limit, provided $\xi_0$ also varies on a
length scale significantly exceeding, $H,$ one can find a  solution of
equation (\ref{cogov}) in the form of an asymptotic series in
ascending powers of $\cs^2$ or equivalently $H^2/|y|^2.$ The zero
order solution being simply $Y=\Phi_\mathrm{Gp} + {\cs^2\op \over
  2\xi_0\Sigma},$ and we note here that the second term is a constant
if $\xi_0$ is constant and may be discarded, or equivalently $\chi=
\Phi_\mathrm{L}.$ The streamlines are then found from (\ref{STRML})
using this value of $\chi.$ We  comment that this result would be
obtained if pressure was neglected completely. Thus we describe it as
the softened gravity limit and it should apply to streamlines at a
distance greater than $\sim H$ from the protoplanet. It can also be
obtained when the approximation of neglecting the acceleration in the
$x$ direction is not made at the outset.  We now consider a
modification of the above model that enables a more realistic
treatment of regions close to the planet. However, this makes it two
dimensional and accordingly more complex. 

\subsection{Modification close to the planet}
We begin by noting that the system may be regarded as being governed
by the components of the equation of motion (\ref{motlocx}) and
(\ref{motlocy}) together with a complete statement of the conservation
of vortensity in the form 
\be \xi_0(\psi)\Sigma = \left(2\op + 
{\partial v_y \over \partial x } -{\partial v_x \over \partial y } 
\right)\label{vort2D}\ee
This equation should be regarded as replacing the continuity equation
in those governing the model. The simple one dimensional model is
obtained by using only (\ref{motlocx}) and (\ref{vort2D}) with $v_x$
being neglected. Equation (\ref{motlocy}) is then used as an auxiliary
equation to subsequently determine  $v_x.$ We now retain $v_x$ in equation (\ref{vort2D})
which now needs to be specified using  equation (\ref{motlocy}) in
advance. We shall retain the approximation of neglecting $v_x$ or
equivalently the radial acceleration  equation in
(\ref{motlocy}). Thus  as before we have 
\be v_y =  { 1\over 2\op}{\partial \chi\over \partial x}.
\label{cost1}\ee
Equation(\ref{motlocx}) gives
\be v_x \left(2\op+{\partial v_y \over \partial x }\right )=
-{\partial (\chi + v_y^2/2)\over  \partial y}  
\label{cost2}.\ee
We make the approximation of replacing $v_y$  in this equation by the
unperturbed Keplerian value $v_y = -3x\op/2.$This leads to 
\be v_x =  -{2\over \op}{\partial \chi \over  \partial y} 
\label{cost3}.\ee
We comment that we  have found the above equation to be satisfied
 close to the planet in numerical  simulations (see below).
 
\noindent Using (\ref{cost1} ), (\ref{cost3}) and  again adopting a
power law equation of state with $\beta =2$ 
and using (\ref{ENTH}), equation (\ref{vort2D}) leads to
\be {\cal D}^2 Y  =
{2\op \xi_0\Sigma \over \cs^2}( Y -\Phi_\mathrm{Gp} )-\op^2 
 \label{2DEQ},\ee
where the operator 
\be {\cal D}^2 \equiv {\partial \over \partial x^2}
+ 4{\partial \over \partial y^2}\ee
and we recall  $Y=\chi +3\op^2x^2/2.$ 
The required solution of equation (\ref{2DEQ}) can be written as 
\be
Y =-\int^{\infty}_{-\infty}
G(x,y, x',y') 
\left(\op^2 +
\left({2\xi_0\Sigma\op \over \cs^2}\right)\Phi_\mathrm{Gp}
\right)dx'dy',\label{GRN2D}\ee 
where $G(x,y, x',y')$ is a two dimensional Green's function. When the
vortensity is constant, and the Green's function that vanishes
at $\infty,$ it is given by 
\be G(x,y, x',y')= -{1\over 4\pi}K_0(k\sqrt{(x-x')^2+(y-y')^2/4}),\ee
where as in standard notation  $K_0$ is the Bessel function (this is
readily obtained after scaling the $y$ coordinate to transform ${\cal
  D}^2$ into the Laplacian). When comparing with the corresponding
result obtained from the  one dimensional  treatment given by equation
(\ref{GRN1D}), we see that there is a two dimensional as compared to a
one dimensional integration. The additional integration over $y'$ acts
like a smoothing on a length scale $H.$ 

The streamlines are in general determined from the Bernoulli condition
that $({\bf v} ^2/2 + w +\Phi_\mathrm{L})$ be constant on them. This
reduces to the condition given through equation (\ref{STRML}) under
the approximation scheme used to derive (\ref{2DEQ}). We further
remark that the scale of the flow in the $y$ direction  then becomes
larger than  $H$ so we may perform the integration over $y'$ keeping
other quantities fixed. Using the result that
\begin{eqnarray}
 {1\over 2}
 \int^{\infty}_{-\infty}K_0(k\sqrt{(x-x')^2+(y-y')^2/4}) dy' =\nonumber \\
 {\pi\over |k|}\exp(-|k(x-x')|),\end{eqnarray}
we then recover equation (\ref{GRN1D}) for the simple one dimensional
 model. 

We may also develop an asymptotic sequence as in section \ref{asymp}
starting with  $Y =\Phi_\mathrm{Gp} -\op^2 +{\cs^2
  \op}/(2\xi_0\Sigma),$ with the last two terms being constant, and
then iterating (\ref{2DEQ}) to find successive corrections to $Y.$
Thus use of the two dimensional model imparts a smoothing in the $y$
direction, with a length scale $H,$ to the one dimensional model. This
causes differences close to the planet for small softening.

\begin{figure*}
\begin{center}
\includegraphics[width= \textwidth]{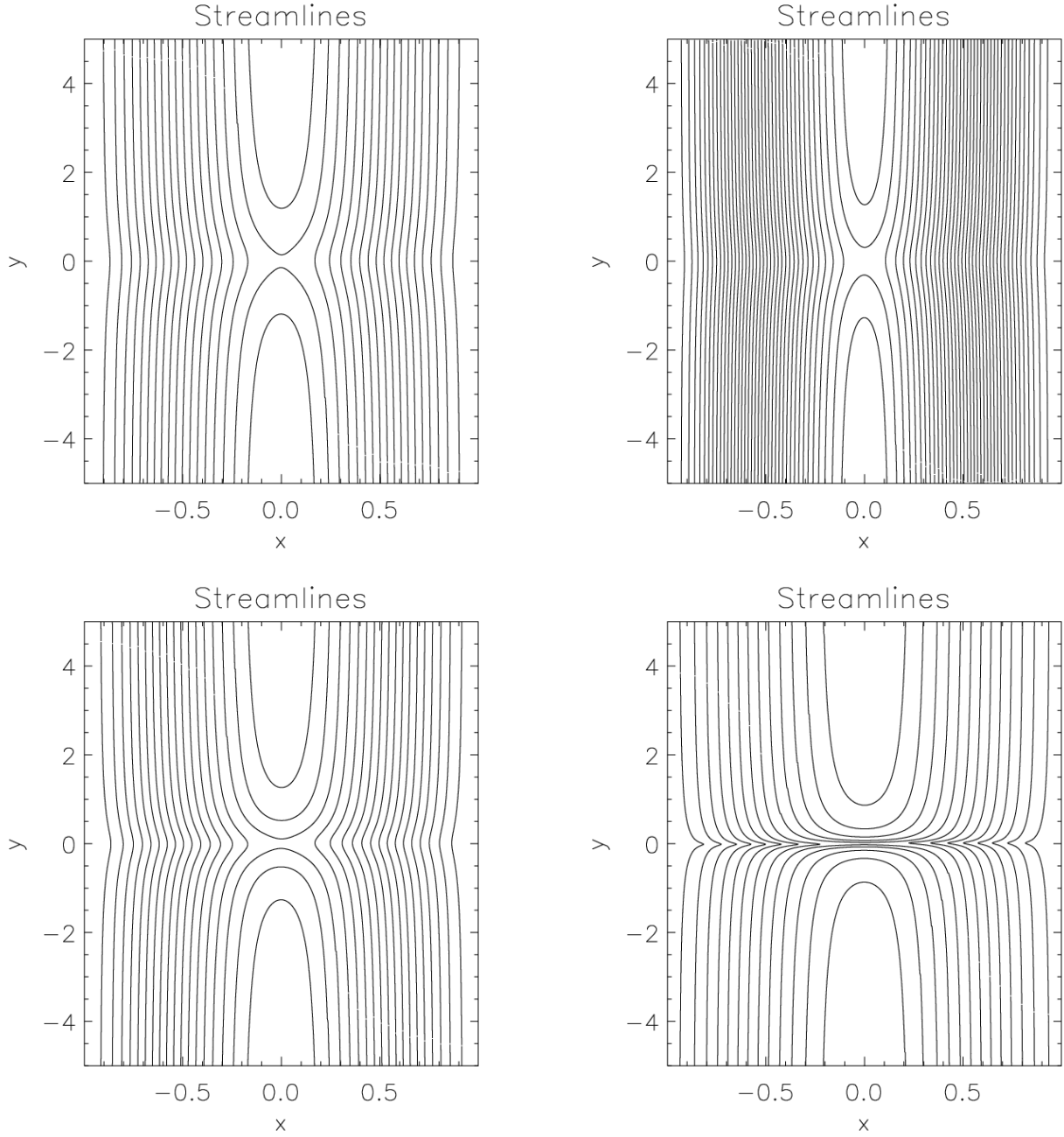}
\caption{  Streamlines for  constant vortensity 
obtained using the simple one dimensional model. The plots are for
  $q/h^3 = 0.0252$
and $b/h = 0.6$ upper left panel, $q/h^3 = 0.0126$  and $b/h = 0.6,$
upper right panel, $q/h^3 = 0.0252$  
and $b/h = 0.3,$ lower left panel and 
 $q/h^3 = 0.0252$ and $b/h = 0.0252,$ corresponding
to the softening parameter being equal to the Bondi radius,
 lower  right panel. For this and other similar figures
the unit of length is the disc scale height, $H,$
evaluated at the origin of the unperturbed flow. }
\label{fig01}
\end{center}
\end{figure*}

\begin{figure*}
\begin{center}
\includegraphics[width= \textwidth]{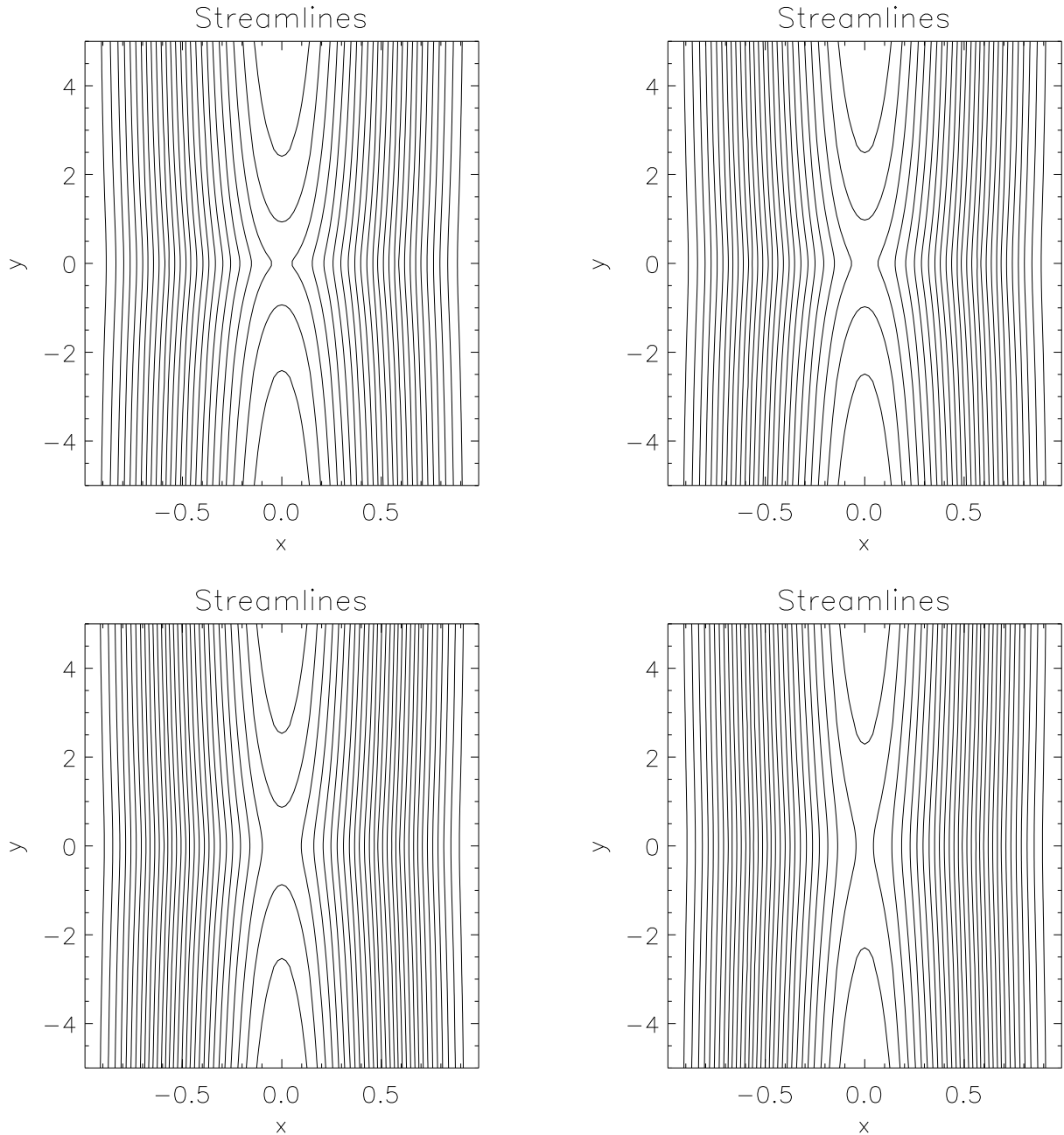}
\caption{ 
 Streamlines for  constant vortensity
 found using the two dimensional Green's
function. The plots are for
  $q/h^3 = 0.0252$
and $b/h = 0.0252,$ corresponding
to the softening parameter being equal to the Bondi radius,
 upper left panel, and $q/h^3 = 0.0252$   with $b/h = 0.1,$
upper right panel.
 In the lower  left
panel,
 $q/h^3 = 0.0252$
 with $b/h = 0.3,$  and in the lower right panel 
 $q/h^3 = 0.0252$  with $b/h = 0.6.$
The horseshoe region is wider for  smaller softening lengths and
converges to a well defined structure for $b \rightarrow 0.$}  
\label{fig02}
\end{center}
\end{figure*}

\begin{figure*}
\begin{center}
\includegraphics[width= \textwidth]{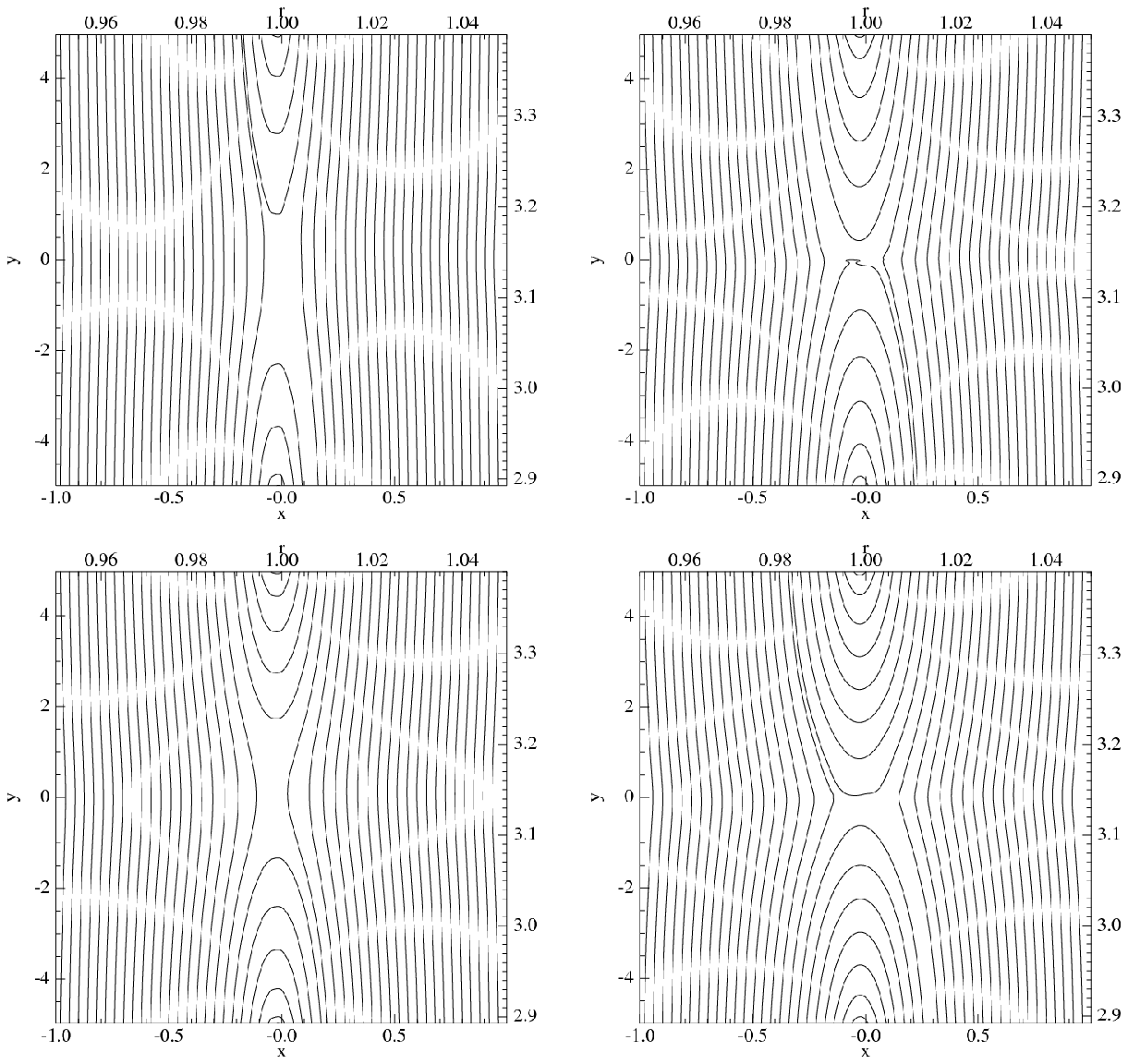}
\caption{Streamlines, obtained from hydrodynamical simulations for constant vortensity and $q/h^3=0.0252$. Left panels: $b/h=0.6$, right panels: $b/h=0.025$. Top panels: full potential
without cut off, bottom panels: the cut off procedure was adopted.}
\label{fig04}
\end{center}
\end{figure*}

\section{Streamline calculations}
\label{secStream}

We have used the simple one dimensional
model to obtain $\chi$ and determine  streamlines
for the  constant vortensity case using a range of softening lengths.
We perform the integral specified in equation (\ref{GRN1D})
in conjunction with equation (\ref{STRML}).
For comparison purposes we also  obtained streamlines  using  equation
(\ref{GRN2D}), instead of (\ref{GRN1D}). In this case the Green's
function and integration are of course two dimensional. 

\subsection{Dimensionless scalings}
Adopting $H$ evaluated at the origin for the unperturbed flow as the
 unit of length, it is straightforward to see that apart from $\xi_0,$
 the distribution of specific vorticity, if $b/h$ is fixed,  the
 streamlines are characterised by only one parameter, $q/h^3,$ where
 $q$ is  ratio of the mass of the protoplanet to the mass of central
 star and $h =H/\rp$ is the disc aspect ratio \citep[see
 e.g.][]{1996ApJS..105..181K}. We work with the power law equation of
 state with  $\beta =2.$ In that case $\cs^2/\Sigma$ is constant. In
 the local model, the adoption of a linear shear means that the
 background vorticity is a constant $=(1/2)\op.$ Therefore for
 constant vortensity the background surface density is a constant
 $=\Sigma_0$ and $\xi_0 = \op/(2\Sigma_0).$ The units are arbitrary
 and so we choose these such that $\Sigma =1$ at the origin in the
 unperturbed flow.   

Streamlines for  constant vortensity  obtained for the simple one
 dimensional model from (\ref{GRN1D}) with  $q/h^3 = 0.0252$
and $b/h = 0.6$ are shown in Fig. \ref{fig01}. This parameterization
 corresponds to $1$ $\me$ in a disc with $H/\rp \sim 0.05.$ The
 streamlines  fall naturally into two groups, the first coming from $y
 >0$ and $x>0$ and the second coming from $y <0$ and $x<0.$ For each
 of these classes a subset passes by the planet while the remainder
 undergoes a horseshoe turn. Those undergoing horseshoe turns
 constitute the horseshoe region, which is separated from the
 remaining domain by two separatrices. For the constant vortensity
 case, the flow has both left-right and up-down symmetry implying an X
 point at the centre  of the protoplanet. When the vortensity is not
 constant this symmetry is in general  lost.

The streamlines shown in the upper left panel of  Fig. \ref{fig01}
corresponding to $q/h^3 = 0.0252$  and $b/h = 0.6.$ have  a horseshoe
width of $\sim 0.25H.$ To illustrate the dependence on $q$ we also
show the streamlines for  $q/h^3 = 0.0126$  and $b/h = 0.6.$ These
suggest that the horseshoe width is proportional to $q^{1/2}.$ This
result can also be derived  by considering the streamline at the centre of the protoplanet. At this location, where $x=
\partial \chi/\partial x =0,$ equation (\ref{GRN2D}) gives 
\be \chi = Y =-{1 \over H^2}\int^{\infty}_{-\infty}
G(0,0, x',y')\Phi_\mathrm{Gp}
dx'dy' +{\cs^2
  \op}/(2\xi_0\Sigma),\label{GRN2D0}\ee
there being a corresponding expression derived from  equation (\ref{GRN1D}).
The horseshoe width, $\xs,$  is obtained by equating this expression to
${\cs^2
  \op}/(2\xi_0\Sigma) - 3\op^2\xs^2/8,$ being the value of $E_0(\psi)$ obtained from (\ref{STRML})
at large distances. Hence very generally,  
\be \xs \propto \sqrt{|\Phi_\mathrm{Gp}|} \propto
\sqrt{q}.\label{WIDTH}\ee 
To be more precise, performing the integral (\ref{GRN2D0}) for the two
dimensional case we find that 
\begin{equation}
\frac{\xs^2}{\rp^2}=\frac{4q}{3\pi h}\int^\infty_0\int^{2\pi}_0 
\frac{K_0(r)r}{\sqrt{r^2(1+3\sin^2\theta)+b^2/h^2}}d\theta dr,
\end{equation}
which can be simplified to
\begin{equation}
\frac{\xs^2}{\rp^2}=\frac{16q}{3\pi h}\int^\infty_0
\frac{K_0(r)r}{\sqrt{4r^2+b^2/h^2}}
E\left(\sqrt{\frac{3r^2}{4r^2+b^2/h^2}}\right) dr, 
\label{eqwidth}
\end{equation}
where $E$ denotes the complete elliptic integral of the first
kind. For $b=0$, this gives 
\begin{equation}
\left.
\xs\right|_{b=0}=\sqrt{\frac{4q}{3h}E\left(\frac{\sqrt{3}}{2}\right)}\rp
\approx 1.68\sqrt{\frac{ q}{h}}~\rp, 
\label{eqxsb0}
\end{equation}
which gives the horseshoe width in the limit of zero softening for the
two dimensional model. We also note that in the limit $b \gg h$, 
\begin{equation}
\xs \sim \sqrt{\frac{8q}{3b}}\rp,
\label{eqxshighb}
\end{equation}
which is the same result that would be obtained by neglecting pressure
effects altogether. 

To investigate the dependence on the softening parameter, we plot streamlines 
for the cases 
 $q/h^3 = 0.0252,$ with
 $b/h = 0.3$  and also with
  $b/h = 0.0252$  in Fig. \ref{fig01}. The second case
corresponds to the softening parameter being equal to the Bondi radius
$\cs^2/(G\mpl)$ for $1$ $\mathrm{M}_{\oplus}$ and a disc aspect ratio
of $H/\rp=0.05.$ Interior to the Bondi radius the gravity of the
planet dominates and we expect a hydrostatic structure that does not
participate in the horseshoe dynamics. We here remark that equation
(\ref{GRN1D}) obtained for the simple one dimensional model diverges
logarithmically for  $ b\rightarrow 0,$ but (\ref{GRN2D}) is
convergent. 
 
In general as expected from (\ref{WIDTH}) the horseshoe width
increases as $b$ decreases and the streamlines become more compressed
and nearly horizontal  near $y=0$ indicating that the acceleration in
the radial direction should not be neglected for small softening as
has been done in the simple one dimensional model. In spite of this,
the behaviour of the streamlines for $y>H$ is very similar for fixed
$q$  and  the different values of $b$ in these cases. This is  in line
with the existence of the asymptotic solution discussed above.

Streamlines for  constant vortensity when the two dimensional Green's
function is used are given in Fig. \ref{fig02}. These are  for
$q/h^3 = 0.0252$ with $b/h = 0.0252,$ corresponding to the softening
parameter being equal to the Bondi radius, together with plots
obtained for   $b/h = 0.1,$  $b/h = 0.3$ and  $b/h = 0.6.$ Use of
the two dimensional Green's function smooths the potential and makes
the horseshoe region narrower compared to the simple one dimensional
case.   When $b=0.6h,$ the horseshoe width was $0.21H$ being about
$20\%$ smaller than that found using the simple one dimensional
model. However, the horseshoe width converges in the limit of zero
softening being about $30\%$ wider than when $b/h =0.6.$

We remark that using a potential vertically averaged with weight $\rho,$
equivalent to a vertical smoothing,  but retaining
the formalism leading to equation(\ref{GRN1D}) has a similar effect
to using the two dimensional Green's function including convergence
for $ b \rightarrow 0.$

\begin{figure}
\centering
\resizebox{\hsize}{!}{\includegraphics[]{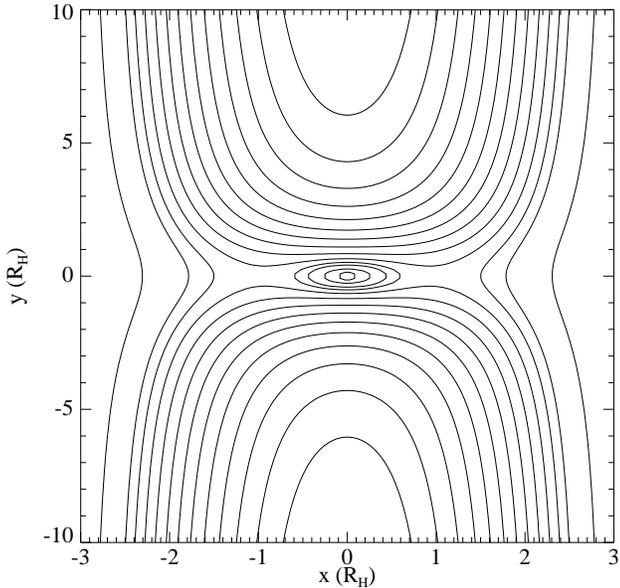}}
\caption{Streamlines close to the planet in the limit $q \gg
  h^3/\sqrt{3}$, in which case a circulating region within the Hill
  sphere arises. The unit of distance in this figure is Hill radius
  $R_H=(q/3)^{1/3}\rp$.}  
\label{figstreamhill}
\end{figure}

The discussion presented above was for low-mass planets, for which
the Bondi radius is smaller than the radius of its Hill sphere $r_p(q/3)^{1/3}.$
 This condition is equivalent to the requirement that 
$q<h^3/\sqrt{3}$. In this regime
 we found no circulating streamlines
close to the planet when the vortensity is constant. 
For more massive planets with $q > h^3/\sqrt{3}$, a circulating region is
expected to interior to  the Hill sphere. This limit was recently
considered analytically in \cite{adamthesis}. We show an example
of the resulting streamlines 
obtained from our model  in this limit in Fig. \ref{figstreamhill}.  Even though
important non linear effects  such as the 
onset of gap formation are not included \citep {2007prpl.conf..655P}
 there is indeed  a
circulating region within the Hill sphere in this case.

However such closed streamlines might in general be expected
 significantly interior to the Bondi radius
for matter bound to
 a low mass planet. We comment that this issue
 is sensitive to model details such as the specification
 of the vortensity profile.
  To illustrate this, consider the situation
  when this bound matter appears almost stationary in the rotating frame
  such that circulating streamlines would be expected.
This matter would have to have significantly lower vortensity
than its surroundings. To see this we note that for a staionary solution
$\chi =0,$ and $Y= 3\Omega_p^2x^2/2.$
With this specification,  equation (\ref{2DEQ}) gives

\be {2\op \cs^2\over \xi_0\Sigma}=
 {3\op^2 x^2 \over 2} -\Phi_\mathrm{Gp}  
 \label{2DEQREV},\ee
 setting  the vortensity profile to be such that  $\xi_0 = 2\op/\Sigma,$
 we obtain
 \be \cs^2 =
 {3\op^2 x^2 \over 2} -\Phi_\mathrm{Gp}  
 \label{2DEQREV1},\ee
 which is the expected condition for hydrostatic equilibrium
 of the protoplanet under centrifugal and tidal forces
 for our baratropic equation of state. Note that to achieve
 this, the surface density should  rapidly increase and hence
 the prescribed vortensity should correspondingly decrease towards the center
 of the protoplanet. On account of the attainment
 of a limiting form of the horseshoe region once the  softening lenght $b$ 
 is smaller than the Bondi radius, we do not expect the details of behaviour at 
 interior radii to significantly affect the results.

\begin{figure}
\centering
\resizebox{\hsize}{!}{\includegraphics[]{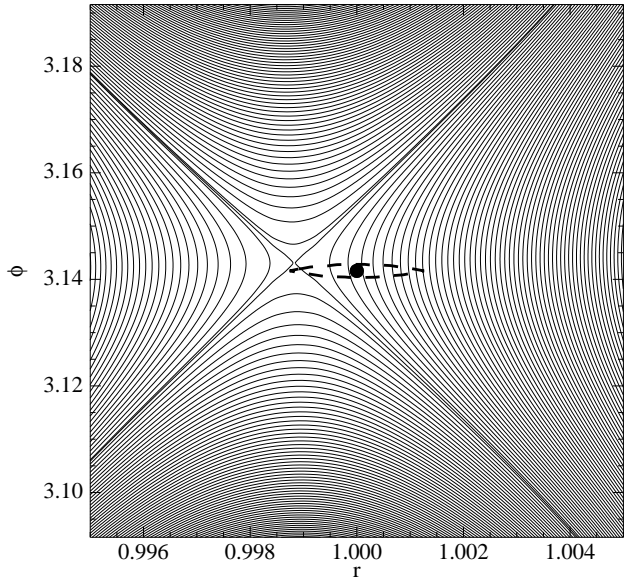}}
\caption{Streamlines close to the planet for $q/h^3=0.0252$ and
  $b/h=0.6$ for a disc with constant vortensity. The planet is
  indicated by the filled circle, and the dashed contour indicates the
  Bondi radius. The cut off procedure was adopted for  this
  simulation.} 
\label{figstreamcut}
\end{figure}

\begin{figure}
\centering
\resizebox{\hsize}{!}{\includegraphics[]{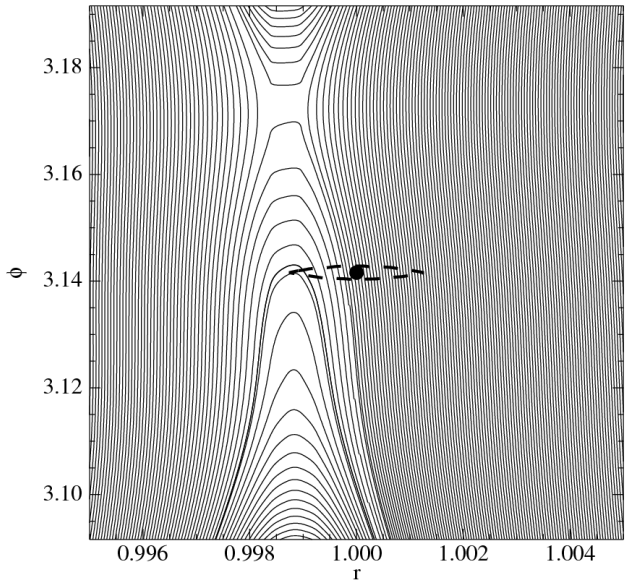}}
\caption{Streamlines close to the planet for $q/h^3=0.0252$ and
  $b/h=0.6$ for a disc with constant vortensity. The planet is
  indicated by the filled circle, and the dashed contour indicates the
  Bondi radius. The full planet potential was adopted with no cut off.}
\label{figstreamnocut}
\end{figure}

\section{Numerical simulations}
\label{secNum}
In this section, we compare the streamline calculations of Section
\ref{secStream} with fully non linear hydrodynamical simulations.  

\subsection{Set up}
We use the RODEO method \citep{2006A&A...450.1203P} in two spatial dimensions, on a regular grid which when at its most
extended, runs from $r=0.5\rp$ to $r=1.8\rp$ and which covers the whole $2\pi$ in azimuth. Since we want to resolve the horseshoe region for even the smallest planets we consider, a relatively high resolution is used.  For the most extended grid
this has $1024$ cells in the radial and $4096$ cells in the azimuthal direction. Then the resolution at the location of the planet is approximately $0.0015\rp$ in both directions. Tests have shown that this resolution is sufficient to capture the horseshoe dynamics for
 $\xs>0.004\rp$. We always ensure that we resolve the softening length $b\rp$ by at least 3 grid cells, which means that for the smallest values of $b$ we consider an even higher resolution is was adopted. We take the disc to be inviscid and isothermal with uniform specific vorticity such that $\Sigma \propto r^{-3/2}$ for the unperturbed initial state. 
 
We consider two kinds of
simulation, the first adopted the perturbing
potential given by equation (\ref{pot}) and adopted
the most extended domain. 
The second type of simulation considers only the coorbital region
extending from $\rp-2H/3$ to $\rp+2H/3$ and occupying
the full $2\pi$ in azimuth. Non reflecting boundary conditions are applied at the radial boundaries such that material
is allowed to leave and enter freely. We refer to this second
type of simulation as having employed a cut off procedure.  Results obtained
with this type of simulation were checked with simulations
employing the most extended domain but modifying
the protoplanet perturbing potential such that it is
 given by equation (\ref{pot}) for $|r-\rp|<2H/3$, and zero otherwise.  These gave very similar results. 
 The cut-offs applied in these 
 simulations  exclude the bulk of  the contributions from the
  Lindblad torques to the flow.  Since either the planet potential vanishes in the region where they are normally generated or that region falls outside the computational domain. As remarked in Section \ref{secCor}, features on the order of one scale height away, such as phenomena associated with Lindblad
  torques,  may affect the coorbital region even when this is much narrower than $H$. Differences between the results 
  obtained from  the two kinds of simulation
we performed are consistent with this supposition.

\begin{figure*}
\centering
\includegraphics[width= \textwidth]{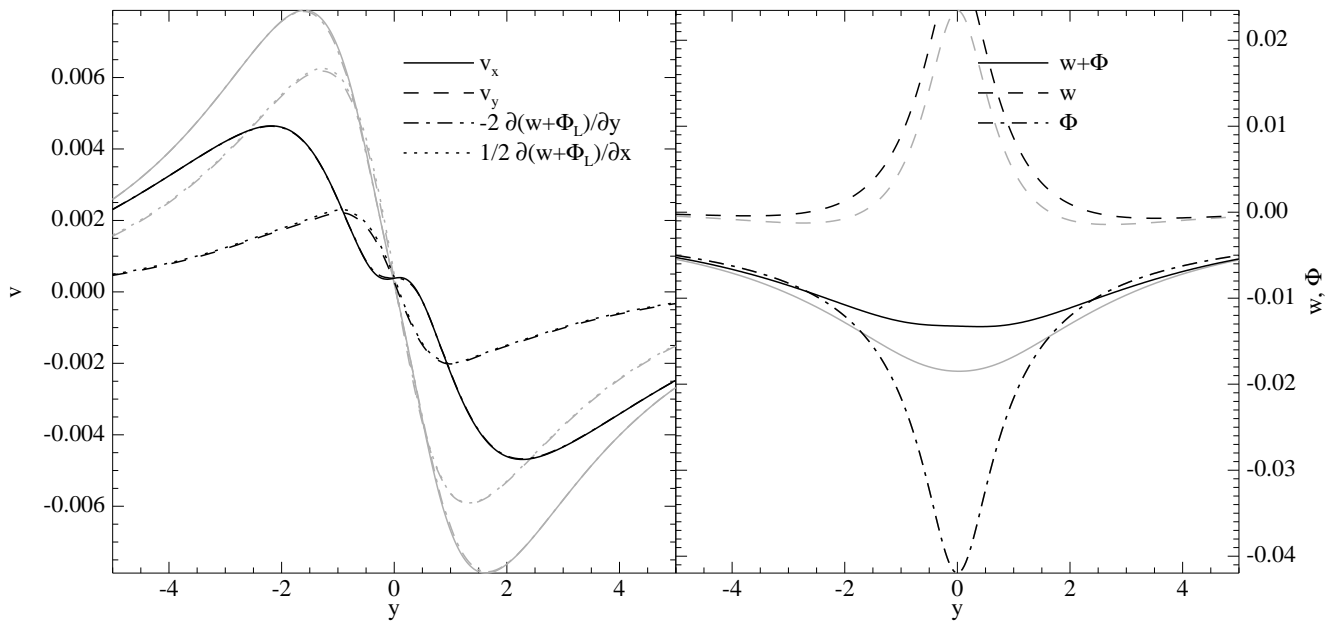}
\caption{Velocities, enthalpy and planet potential at the radial location of the stagnation point for $q/h^3=0.0252$ and $b/h=0.6$. Black curves indicate that the full planet potential was used, grey curves indicate results obtained with the cut-off procedure adopted.}
\label{figvx_dhdy}
\end{figure*}

\subsection{Streamline analysis}
We start by comparing the flow on a scale of $H$ to that found
using the model of Section \ref{secCor} in Fig. \ref{fig04}, for two different values of the softening parameter $b$ and the two 
types of simulation. To obtain $q/h^3=0.0252$ we set up
1 $\me$ around a Solar mass star in a disc with $h=0.05.$ We compare the results illustrated in this figure with the corresponding  results obtained using  the two dimensional Green's function shown in Fig. \ref{fig02}. For the case of large softening, $b/h=0.6$, without a cut off,  although the streamline pattern appears similar, the width of the horseshoe region is about $22\%$ smaller than that depicted in Fig. \ref{fig02}.
 When the cut off procedure is applied,
  the horseshoe region increases in width by about $50\%$
 (see below).  This indicates that phenomena located in the region where the Lindblad torques are generated may play a significant role in shaping the coorbital region.

For smaller softening such that $b/h=0.025$, corresponding to the softening parameter being equal to the Bondi radius, the 
differences between the two types of simulation  is  less extreme. The measured increase in $\xs$ 
obtained on applying the cut off procedure  amounts to $20 \%.$
In this case there is very good agreement
between the simulation without a cut off and
 the model of Section \ref{secCor} with the two dimensional Green's function illustrated in Fig. \ref{fig02}, with the values of $\xs$ differing by several percent.

  The dependence on softening can be understood as follows: the strength of phenomena related to Lindblad torques is determined by the planet potential at approximately a distance $H$ from the planet, while the width of the horseshoe region depends on $|\Phi_\mathrm{Gp}|$  at the location of the planet. For a softening parameter comparable to $h$, the planet potential at these two locations is comparable. Therefore, it is relatively easy for  Lindblad torque related phenomena  to affect the coorbital region. For smaller softening, $|\Phi_\mathrm{Gp}|$ at the location of the planet  increases, while the value at a distance $H$ remains largely unchanged. For $b\ll h$, we expect the effect of phenomena originating a distance $\sim H$ from the planet to be smaller, and therefore there should be  better agreement between the numerical simulations and  the model of Section \ref{secCor}.

The effect of introducing the cut off is to produce larger corotational speeds directed towards the planet. These are eventually slowed down as the pressure gradient reduces, producing a stagnation point at, or very close to the planet's location. The faster moving material can originate further from the planet and so is associated with an increased horseshoe width. 

The weaker corotational flow that occurs without the cut off is associated with significantly increased pressure in the region near $\varphi=\pp$. This back pressure is affected by conditions a distance $\sim H$ from the planet and it can  distort the streamline pattern close to the planet so that it becomes asymmetric as shown in Figs. \ref{figstreamcut} and \ref{figstreamnocut}. For the larger softening cases, the stagnation point is displaced azimuthally a distance  $\sim b\rp$ from the location of the planet and the  horseshoe width is  reduced (see Fig. \ref{fig04}).    

However, when the cut-off is applied, there is a single stagnation point at $\varphi=\pp$, slightly displaced from the radial location of the planet due to the radial pressure gradient of the unperturbed disc.
The latter  is a very minor effect that is absent from the
local models on account of their strict  symmetry.
The local models always produce a single stagnation point at
the location of the planet, a situation that is essentially recovered 
for small softening in simulations without a cut off (see also Fig. \ref{fig04}). 

We take a closer look at this back pressure effect in Figs. \ref{figvx_dhdy} and \ref{figdhdy_b0,03}, where we consider a slice through $r=r_\mathrm{stag}$, where $r_\mathrm{stag}$ is the radial location of the stagnation point. Because of the radial pressure gradient of the unperturbed disc, $r_\mathrm{stag} \ne \rp$ (see Figs. \ref{figstreamcut} and \ref{figstreamnocut}). From equations (\ref{cost1}) and (\ref{cost3}) we know that the velocity is directly related to the gradient of $\chi$. First of all, we can check if this approximation is valid. From the left panel of Fig. \ref{figvx_dhdy}, we see that for both the full and the cut-off potential there is very good agreement between both $v_x$ and $v_y$ as measured from the simulation and the
values expected from equations (\ref{cost1}) and (\ref{cost3}).  Therefore, their assumption in the two dimensional model
of section \ref{secCor}  
should be  a good approximation. 

Note, however, the strong differences between the black curves for the full potential and the grey curves for the cut-off potential. Velocities are reduced due the back pressure originating at distances $\sim H$ from the planet when there
is no cut off. The right panel of Fig. \ref{figvx_dhdy} clearly shows  a higher peak in the enthalpy for the  potential without a cut off. When added to the protoplanet potential
the total is less in magnitude in that case and thus we can see from equations (\ref{cost1})
and (\ref{cost3}) that smaller  inflow velocities will accordingly be produced. 
 
\begin{figure}
\centering
\resizebox{\hsize}{!}{\includegraphics[]{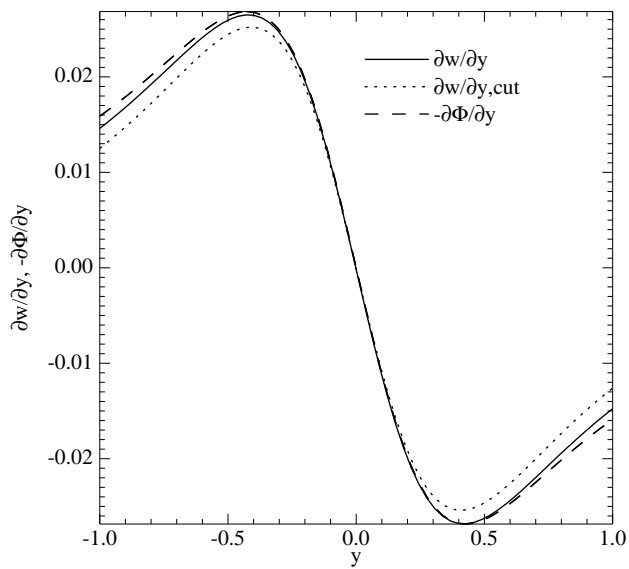}}
\caption{Derivatives of the enthalpy and the potential at the orbital radius of the stagnation point for $q/h^3=0.0252$ and $b/h=0.6$.}
\label{figdhdy_b0,03}
\end{figure}

\begin{figure}
\centering
\resizebox{\hsize}{!}{\includegraphics[]{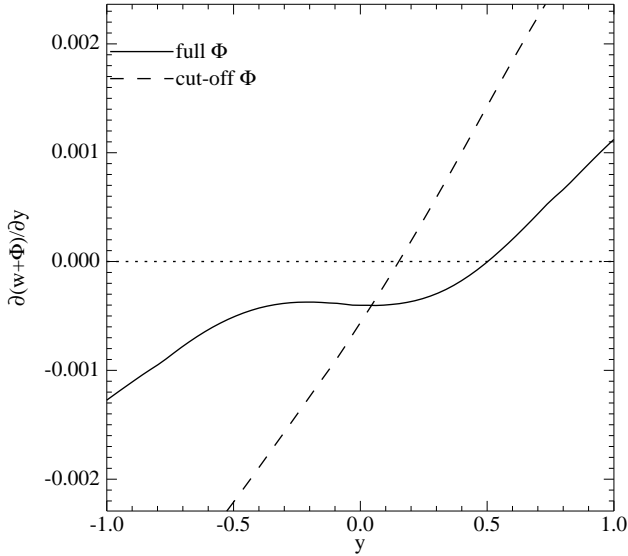}}
\caption{Derivative of the sum of the enthalpy and the potential at the orbital radius of the stagnation point for $q/h^3=0.0252$ and $b/h=0.6$, with and without the cut off procedure.}
\label{figwplusphi1}
\end{figure}

The azimuthal shift of the stagnation point can be explained as follows. At a stagnation point, we must have that $\partial w/\partial y = -\partial \Phi_\mathrm{L}/\partial y$ (see equation (\ref{motlocy})), since all velocities must vanish. When the cut-off is applied, there is only one stagnation point possible (see Fig. \ref{figdhdy_b0,03}) at the azimuth of the planet. This is necessarily the case for a local model in the case of constant specific vorticity, because of symmetry arguments.  Effects originating  a distance $\sim H$
from the planet such as the production of the  Lindblad wakes can destroy this symmetry by means of a back pressure effect. This increases the gradient of $w$ near the planet, which, if strong enough, can give rise to three possible stagnation points. For the case shown in Fig. \ref{figdhdy_b0,03}, the back pressure is such that one stagnation point appears, but shifted in azimuth by approximately one softening length. In Fig. \ref{figwplusphi1}, we show the $y$-derivative of $w+\Phi_\mathrm{L}$ at the radial location of the stagnation point. 

 When no cut off is used, the back pressure gives rise to a gradient of $w$ at $y=0$, pushing the stagnation point away from the planet. The actual configuration of the stagnation points depends on details in the flow, for example the background surface density gradient. Indeed, \cite{2006ApJ...652..730M} reported three stagnation points for the case with constant background surface density, a configuration that we find as well for the same surface  density gradient. Since both the back pressure and the potential are proportional to $q$ in the linear regime, the location of the stagnation points does not depend on the planet mass, which was also reported by \cite{2006ApJ...652..730M}.  
   
When $b\ll h$, the Lindblad torques will be largely independent of the softening parameter, since they are generated at distances much larger than $b\rp$ in this case. We then expect any related back pressure to be small, with any stagnation points located close to the planet. We find this indeed to be the case, and we  see that for $b\ll h$ the agreement of $\xs$ as measured from the simulations with that found using the model of Section \ref{secCor}, which does not take account of phenomena
related to epicyclic motions such as Lindblad torques, is much better. 
 
\begin{figure}
\centering
\resizebox{\hsize}{!}{\includegraphics[]{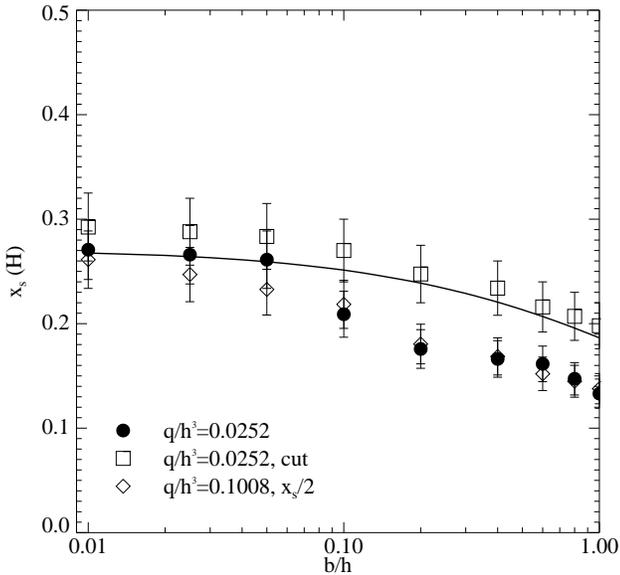}}
\caption{Half-width of the horseshoe region vs. softening parameter. Two planet masses are considered, the lowest one with and without the cut-off procedure. For the more massive planet $\xs/2$ is shown, to remove the $\sqrt{q}$ dependence. Error bars indicate an error of $10\%$.}
\label{figwvsb}
\end{figure}

\subsection{Width of the horseshoe region}
We have measured the half-width of the horseshoe region $\xs$ for various values of $b$ and $q$ though a streamline analysis. The results are shown in Fig. \ref{figwvsb}, with $10\%$ error bars indicating our error estimate, for the cases with and without 
applying the cut off procedure. As indicated in our
above discussion, horseshoe widths are always larger when the 
cut off procedure is applied. The deviation varying from about $10\%$ at very small softening to $50 \%$ when $b \sim h.$
  The results predicted by the model in Section \ref{secStream} with the two-dimensional Green's function are indicated by the solid curve. These are in good agreement
  with the other results for small softening but give values for $\xs \sim 20\%$
  larger than those found from the simulations without a cut off procedure.
It appears that  these results  fall below the others, because of the effect
discussed above that we described as being due to a back pressure related to phenomena
such as the wake produced at a distance $\sim 2H/3$ from the planet.
 As expected, for smaller values of $b$ this effect is reduced. 

In Fig. \ref{figwvsb} we also show results for a planet with a mass that is four times larger (corresponding to 4 $\me$ orbiting a  Solar mass star, embedded in a disc with $h=0.05$).
   For this case, the measured value of $\xs$ was divided by 2 
   to remove the $\sqrt{q}$ scaling.
    If $\xs\propto \sqrt{q}$ in this mass range,
     the black circles and diamonds should fall on top of each other 
     in Fig. \ref{figwvsb}.
      It is clear that for all values of the softening parameter
       that we consider, the horseshoe width scales as $\sqrt{q}$ 
       in this mass range. 
       \cite{2006ApJ...652..730M} speculated that this scaling would brake 
       down for softening parameters smaller
        than the Bondi radius.
        We find this not to be the case, since  
        our smallest softening parameter $b<q/h^2$ for both planet masses
        placing it in that regime.
        
         Although the perturbed surface density at the location
          of the planet can be quite large for small softening,
           this perturbation is almost in hydrostatic equilibrium and so 
           does not play 
           a major role in the flow, and does not cause a departure from $\xs\propto \sqrt{q}$. 
             Below, we will argue that instead, 
             while the value of $b$ is important,
              this departure is governed by the ratio $q/h^3.$ 

Additional models with different values of $H,$
 while keeping $b/h$ fixed, confirm that $\xs \propto 1/\sqrt{h}$ (see equation (\ref{eqwidth})) in the same mass range, as long as $\xs \ll H$.

\subsection{Horseshoe drag}
\citet{1991LPI....22.1463W} found an expression for the corotation
torque produced by material in the coorbital region in the form
\begin{equation}
\label{HDRAG}
\Gamma_\mathrm{c,hs}=\frac{3}{4}\left(\frac{3}{2}-\alpha\right)\xs^4 \Sigma
\rp^4\op^2,
\end{equation}
where the surface density $\Sigma \propto r^{-\alpha},$ and all quantities
are evaluated at $r=\rp.$
On  the other hand \cite{2002ApJ...565.1257T}
found an expression for the corotation torque derived from 
linear perturbation analysis in the form
\begin{equation}
\label{eqTlin}
\Gamma_\mathrm{c,lin}=1.36\left(\frac{3}{2}-
\alpha\right)\frac{q^2}{h^2}\Sigma \rp^4\op^2.  \end{equation}
Here we shall make use of these expressions, referring 
the reader to a companion paper for additional discussion.

We first note that although they are both torques, these expressions
were derived for systems with differing flow topology and so
should not be expected to be the same. The horseshoe drag applies
to a system with the flow topology of our solutions for the coorbital region
which has separatrices. On the other hand the linear corotation torque
 is derived from the linear perturbation theory of circular orbits.
 
  The torque in both expressions is proportional to the
  vortensity gradient which cancels out when they are equated.
  In this paper we considered only constant vortensity for which there 
  is no torque. But we can consider the case of a small and smooth
   vortensity gradient, and  following the discussion of 
   section \ref{secCor} we can argue, as has also been confirmed
   in simulations, that the horseshoe width should be close to that found
   assuming constant vortensity. 
   
\cite{2006ApJ...652..730M} assumed that  torques obtained from
simulation results with $b/h=0.3$ could be used to determine values
of $\xs$ by equating them to torques obtained from
\cite{2002ApJ...565.1257T} even though the latter were calculated
for $b=0.$ The determined values of $\xs$ agreed with those directly
measured from the simulations, and we remark that our measurements
of $\xs$ for  $b/h=0.3$ agree with those of \cite{2006ApJ...652..730M}.
   
However, from Fig. \ref{figwvsb}, we see that $\xs$ increases by a
factor of $1.67$ as  $b$  decreases from $0.3h$ to zero. Therefore,
the horseshoe drag torque, being proportional to $\xs^4$, is actually
nearly an order of magnitude larger than the linear corotation torque
for $b=0$ and so {\it  should not}  be equated to it. It thus turns
out that  the combined effects of finite softening and the back
pressure phenomenon discussed above reduced the estimated horseshoe
drag by almost an order of magnitude compared to the value that should
have been adopted to compare with  \cite{2002ApJ...565.1257T}.

\begin{figure}
\centering
\resizebox{\hsize}{!}{\includegraphics[]{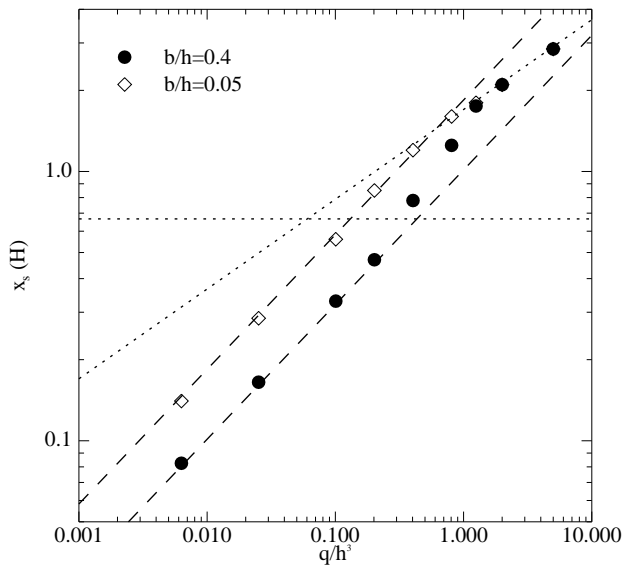}}
\caption{Half-width of the horseshoe region vs. mass ratio $q$, for two different values of $b/h$. The dashed lines indicate $\left. \xs\right|_{b=0}$, given by equation (\ref{eqxsb0}), and $0.6$ times this value to account for the effects of softening and back pressure. The horizontal dotted line indicates $\xs=2H/3$. The tilted dotted line shows equation (\ref{eqxshighq}), expected to be valid at high masses.  }
\label{figwvsq}
\end{figure}

\subsection{Extension to higher masses}

When $q/h^3$ is of order unity, the Hill sphere, the Bondi radius and
the half-width of the horseshoe region are all comparable to $H$. At
approximately this mass, the waves excited by the planet start to
become non-linear at a distance $\sim H$ from it, and gap formation
sets in. This reduces the strength of the Lindblad torques and any
influence of material at a distance $\sim H,$ and therefore the back
pressure effect described above. At the same time, the width of the
horseshoe region,  being proportional to $\sqrt{q/h}$, also becomes
comparable to $H,$ extending  into the Lindblad resonance region. This
also reduces the back pressure effect, making the horseshoe width
larger. This was checked by running an additional model with
$q/h^3=0.4032$ ($8$ $\me$ embedded in a disc with $h=0.05$ around a
Solar mass star), $b/h=1$ and $b/h=0.2.$ For $b/h=1$, we find a
horseshoe width of $\xs=0.54H,$ which corresponds to an exact scaling
$\xs \propto \sqrt{q}$ according to Fig. \ref{figwvsb}. This is to be
expected, since $\xs<2H/3$, and the horseshoe region does not extend
into the wave excitation zone. For $b/h=0.2,$, we find $\xs=0.9H$,
while a scaling with $\sqrt{q}$ would imply $\xs=0.75H$, according to
Fig. \ref{figwvsb}. Therefore, the horseshoe width is larger than
predicted by the simple model, in agreement with the findings of
\cite{2006ApJ...652..730M}. We stress that this behaviour should not
be seen as an onset of nonlinearity, since there is \emph{no}
horseshoe region in linear theory, but rather as a reduction of the
back pressure effect of material at a distance $\sim 2H/3$ from the
planet where the  main Lindblad torques are produced on the horseshoe
region.   

The exact mass at which $\xs=2H/3$ of course  depends on softening. 
 If we write $\xs=C(b)\xs(b=0)$, with $C(b)<1$ and for an approximate
 estimate, set $\xs(b=0)/\rp= 1.68\sqrt{q/h}$ (see equation (\ref{eqxsb0})), we get:
\begin{equation}
q<q_\mathrm{trans}\equiv \frac{0.157 h^3}{ C(b)^2}
\end{equation}
as a condition for $\xs \propto \sqrt{q}$. For $b/h=0.3$, as used by \cite{2006ApJ...652..730M}, we have $C\approx 0.6$ (see Fig. \ref{figwvsb}), and therefore $q_\mathrm{trans}=5.5\cdot 10^{-5}$ for $h=0.05$, exactly the mass at which the departure from linearity as reported by \cite{2006ApJ...652..730M} begins. 
Note that for the same parameters, the ratio of the Bondi radius to the softening parameter is approximately unity. However, for $C=1$, we have $q_\mathrm{trans}/h^3=0.157$, confirming that  for $q/h^3=0.1008$ and $q/h^3=0.0252$, as shown in Fig. \ref{figwvsb}, $\xs\propto\sqrt{q}$ for all softening parameters,  independent of the ratio of the Bondi radius to the softening.
 On the other hand, for $b \gg h$, using equation (\ref{eqxshighb}), it can be shown that $q_\mathrm{trans}/h^2 \propto b$. In this case, the critical value of $b$ for which $q=q_\mathrm{trans}$ is proportional to the Bondi radius. However, such a large softening is incompatible with the idea of vertical averaging, from which we expect $b$ to be of the order of $h$.

These ideas are further illustrated in Fig. \ref{figwvsq}, where we show the measured half-width of the horseshoe region versus planetary mass. For $q \ll h^3$, we expect $\xs \propto \sqrt{q}$, with a coefficient that depends on the softening. For small softening, we see that the results agree very well with equation (\ref{eqxsb0}), while for large softening we find values that are a factor of $0.6$ smaller due to the combined effect of non-zero softening and back pressure (see above). At higher masses, we expect $\xs$ to be proportional to the radius of the Hill sphere \citep{adamthesis}:
\begin{equation}
\xs=2.47 \left(\frac{q}{3}\right)^{1/3},
\label{eqxshighq}
\end{equation} 
indicated by the tilted dotted line in Fig. \ref{figwvsq}.  For $q \gg h^3$ we find this indeed to be the case, independent of the value of $b/h$. The results for $b/h=0.4$ are in very good agreement with \cite{2006ApJ...652..730M}, who used $b/h=0.3$, for all values of $q$.  In between the two regimes of $\xs \propto \sqrt{q}$ and $\xs \propto q^{1/3}$ the width of the horseshoe region rises faster than $\sqrt{q}$, which is due to a reduction in the strength of the back pressure, as argued above. For $b/h=0.05$, no such behaviour is found, since the effects of the back pressure are small for all $q$.

We therefore conclude that for $q>q_\mathrm{trans}$, 
the back pressure effect of the Lindblad torques is reduced,
 which leads to an increase in the width of the horseshoe region
towards the value obtained from equation (\ref{eqwidth}).
 This is consistent with the streamline analysis presented
 in \cite{2006ApJ...652..730M}, where it is shown that in this transition 
regime only one stagnation point survives and moves towards the location of the planet.
 We have confirmed this behaviour in our simulations.
 This increase in $\xs$  can have a major impact
 on the torque acting on the planet (see equation (\ref{HDRAG})).  


\section{Discussion and conclusions}
\label{secDisc}
In this paper we presented a  simple model of the coorbital region around a low mass planet.
Using this we derived the horseshoe width as a function of planet mass and gravitational softening
parameter. In the limit of zero softening we found that
\begin{equation}
\frac{\xs}{\rp}=1.68\sqrt{\frac{q}{h}}.
\label{eqxstan}
\end{equation}
This result agreed with high resolution numerical simulations to within several percent. However
for softening lengths $b\rp \sim H,$ the discrepancy was larger, with the simulations
indicating a horseshoe width about $22\%$ smaller.
By considering simulations for which a cut off procedure was used
to remove the effects of the protoplanet potential produced at and beyond a radial separation of $2/3H$
from it, it was found that phenomena at that separation could significantly affect
the horseshoe width, even when that was much narrower, distorting the streamlines and 
 reducing  the width through the action of an additional back pressure that is more effective
for larger softening. This may artificially reduce the horseshoe drag in such cases.

We  also used our results to show that
 the horseshoe drag, exerted by material executing horseshoe turns
is about an order of magnitude larger than the linear corotation torque in the zero softening limit. 
 A more complete  comparison between linear corotation torques
 and horseshoe drag for finite $b$  requires additional linear calculations which  
are presented in detail in an accompanying paper. 
  There we also find that the non-linear corotation torque (horseshoe drag) 
is always much larger than the linear corotation torque for non zero $b$. 
 
We have focused on a two-dimensional description of the horseshoe region, with a softening parameter $b$ 
in the planet potential which may
 approximately account for three-dimensional effects. 
As reported in \cite{2006ApJ...652..730M}, the  horseshoe drag  torque 
appears to be stronger in fully three-dimensional simulations compared to two-dimensional 
runs that include softening. 
Clearly, a three-dimensional model of the horseshoe region is desirable. 
This will be the subject of a future investigation. 

Another useful extension of the present discussion would be the inclusion of non-barotropic effects. 
We remark that the model presented in this paper is valid for discs that have a
 constant specific vorticity and entropy, the latter condition leading to a barotropic equation of state.
 Introducing a radial vortensity gradient breaks the up-down symmetry in Figs. 1-3,
 but simulations show that this effect is barely detectable. Thus we may
 expect that the prediction of the horseshoe width obtained from our   simple model
  may  work reasonably  for non-barotropic discs with a radial entropy gradient. 
However, we do expect  some difference in $\xs$ between isothermal and non-isothermal discs. 
It is easy to see that $\xs$, as obtained from equation (\ref{GRN2D0}), is proportional to $\cs^{-1/2}$. 
For equal temperatures, the sound speed in an adiabatic disc is a factor $\sqrt{\gamma}$ larger than the
 isothermal sound speed, where $\gamma$ is the adiabatic exponent. 
This makes the horseshoe region a factor $\gamma^{1/4}$ smaller in adiabatic simulations. 
Although the difference lies within our $10\%$ error bars, we have noticed it when
  comparing our present results with those in \cite{2008A&A...485..877P}. 
Note that this makes the adiabatic horseshoe drag, being proportional to $\xs^4$, a factor
 $\gamma$ smaller than the isothermal horseshoe drag. 
Since also the wave torque scales as $\gamma^{-1}$,
 the relative strength of the Lindblad torques and horseshoe drag remains the same for adiabatic discs.  

The shape of the horseshoe region changes when a global radial mass flow is introduced with respect to the planet.
 This mass flow can be due to viscous accretion, but also due to radial movement of the planet
 when allowing the orbit of the planet to change.
 When the time scale of the radial flow with respect to the planet to cross
 the horseshoe region is smaller than the libration time scale,
 an asymmetry between the sides  of the 
horseshoe region leading and trailing the planet develops  \citep{2004ASPC..324...39A}.
 This is important for studying Type III migration \citep{2003ApJ...588..494M}. 
In this paper, we have kept the planet on a fixed orbit in an inviscid disc,
 and therefore  such effects  did not occur. 
More work is necessary to study the importance of including the effect
of planetary migration on the disc response and torques  for low-mass planets.
 
\section*{Acknowledgements}
We thank the anonymous referee for an insightful report. This work was performed using the Darwin Supercomputer of the University of Cambridge High Performance Computing Service (http://www.hpc.cam.ac.uk), provided by Dell Inc. using Strategic Research Infrastructure Funding from the Higher Education Funding Council for England.

\bibliography{paardekooper.bib}

\begin{thebibliography}{}

\bibitem[\protect\citeauthoryear{{Artymowicz}}{{Artymowicz}}{2004}]{2004ASPC..%
324...39A}
{Artymowicz} P.,  2004, in {Caroff} L.,  {Moon} L.~J.,  {Backman} D.,
  {Praton} E.,  eds, ASP Conf. Ser. 324: Debris Disks and the Formation of
  Planets {Dynamics of Gaseous Disks with Planets}.
p.~39

\bibitem[\protect\citeauthoryear{{Baruteau} \& {Masset}}{{Baruteau} \&
  {Masset}}{2008}]{2008ApJ...672.1054B}
{Baruteau} C.,  {Masset} F.,  2008, \apj, 672, 1054

\bibitem[\protect\citeauthoryear{{Crida} \& {Morbidelli}}{{Crida} \&
  {Morbidelli}}{2007}]{2007MNRAS.377.1324C}
{Crida} A.,  {Morbidelli} A.,  2007, \mnras, 377, 1324

\bibitem[\protect\citeauthoryear{{D'Angelo}, {Henning} \& {Kley}}{{D'Angelo}
  et~al.}{2002}]{2002A&A...385..647D}
{D'Angelo} G.,  {Henning} T.,    {Kley} W.,  2002, \aap, 385, 647

\bibitem[\protect\citeauthoryear{{D'Angelo}, {Kley} \& {Henning}}{{D'Angelo}
  et~al.}{2003}]{2003ApJ...586..540D}
{D'Angelo} G.,  {Kley} W.,    {Henning} T.,  2003, \apj, 586, 540

\bibitem[\protect\citeauthoryear{{Goldreich} \& {Tremaine}}{{Goldreich} \&
  {Tremaine}}{1979}]{1979ApJ...233..857G}
{Goldreich} P.,  {Tremaine} S.,  1979, \apj, 233, 857

\bibitem[\protect\citeauthoryear{{Goldreich} \& {Tremaine}}{{Goldreich} \&
  {Tremaine}}{1980}]{1980ApJ...241..425G}
{Goldreich} P.,  {Tremaine} S.,  1980, \apj, 241, 425

\bibitem[\protect\citeauthoryear{{Korycansky} \& {Papaloizou}}{{Korycansky} \&
  {Papaloizou}}{1996}]{1996ApJS..105..181K}
{Korycansky} D.~G.,  {Papaloizou} J.~C.~B.,  1996, \apjs, 105, 181

\bibitem[\protect\citeauthoryear{{Korycansky} \& {Pollack}}{{Korycansky} \&
  {Pollack}}{1993}]{1993Icar..102..150K}
{Korycansky} D.~G.,  {Pollack} J.~B.,  1993, Icarus, 102, 150

\bibitem[\protect\citeauthoryear{{Lin} \& {Papaloizou}}{{Lin} \&
  {Papaloizou}}{1986}]{1986ApJ...309..846L}
{Lin} D.~N.~C.,  {Papaloizou} J.,  1986, \apj, 309, 846

\bibitem[\protect\citeauthoryear{{Masset}, {D'Angelo} \& {Kley}}{{Masset}
  et~al.}{2006}]{2006ApJ...652..730M}
{Masset} F.~S.,  {D'Angelo} G.,    {Kley} W.,  2006, \apj, 652, 730

\bibitem[\protect\citeauthoryear{{Masset} \& {Papaloizou}}{{Masset} \&
  {Papaloizou}}{2003}]{2003ApJ...588..494M}
{Masset} F.~S.,  {Papaloizou} J.~C.~B.,  2003, \apj, 588, 494

\bibitem[\protect\citeauthoryear{{Mayor} \& {Queloz}}{{Mayor} \&
  {Queloz}}{1995}]{1995Natur.378..355M}
{Mayor} M.,  {Queloz} D.,  1995, \nat, 378, 355

\bibitem[\protect\citeauthoryear{{Nelson} \& {Papaloizou}}{{Nelson} \&
  {Papaloizou}}{2004}]{2004MNRAS.350...849N}
{Nelson} R.~P.,  {Papaloizou} J.~C.~B.,  2004, \mnras, 350, 849

\bibitem[\protect\citeauthoryear{{Ogilvie} \& {Lubow}}{{Ogilvie} \&
  {Lubow}}{2002}]{2002MNRAS.330..950O}
{Ogilvie} G.~I.,  {Lubow} S.~H.,  2002, \mnras, 330, 950

\bibitem[\protect\citeauthoryear{{Paardekooper} \& {Mellema}}{{Paardekooper} \&
  {Mellema}}{2006a}]{2006A&A...459L..17P}
{Paardekooper} S.-J.,  {Mellema} G.,  2006a, \aap, 459, L17

\bibitem[\protect\citeauthoryear{{Paardekooper} \& {Mellema}}{{Paardekooper} \&
  {Mellema}}{2006b}]{2006A&A...450.1203P}
{Paardekooper} S.-J.,  {Mellema} G.,  2006b, \aap, 450, 1203

\bibitem[\protect\citeauthoryear{{Paardekooper} \& {Mellema}}{{Paardekooper} \&
  {Mellema}}{2008}]{2008A&A...478..245P}
{Paardekooper} S.-J.,  {Mellema} G.,  2008, \aap, 478, 245

\bibitem[\protect\citeauthoryear{{Paardekooper} \& {Papaloizou}}{{Paardekooper}
  \& {Papaloizou}}{2008}]{2008A&A...485..877P}
{Paardekooper} S.-J.,  {Papaloizou} J.~C.~B.,  2008, \aap, 485, 877

\bibitem[\protect\citeauthoryear{{Papaloizou}, {Nelson}, {Kley}, {Masset} \&
  {Artymowicz}}{{Papaloizou} et~al.}{2007}]{2007prpl.conf..655P}
{Papaloizou} J.~C.~B.,  {Nelson} R.~P.,  {Kley} W.,  {Masset} F.~S.,
  {Artymowicz} P.,  2007, in Protostars and Planets V {Disk-Planet Interactions
  During Planet Formation}.
pp 655--688

\bibitem[\protect\citeauthoryear{{Pepli{\'n}ski}}{{Pepli{\'n}ski}}{2008}]{adam%
thesis}
{Pepli{\'n}ski} A.,  2008, PhD thesis, Department of Astronomy, Stockholm
  University, Stockholm, Sweden

\bibitem[\protect\citeauthoryear{{Pepli{\'n}ski}, {Artymowicz} \&
  {Mellema}}{{Pepli{\'n}ski} et~al.}{2008a}]{2008MNRAS.386..179P}
{Pepli{\'n}ski} A.,  {Artymowicz} P.,    {Mellema} G.,  2008a, \mnras, 386, 179

\bibitem[\protect\citeauthoryear{{Pepli{\'n}ski}, {Artymowicz} \&
  {Mellema}}{{Pepli{\'n}ski} et~al.}{2008b}]{2008MNRAS.387.1063P}
{Pepli{\'n}ski} A.,  {Artymowicz} P.,    {Mellema} G.,  2008b, \mnras, 387,
  1063

\bibitem[\protect\citeauthoryear{{Tanaka}, {Takeuchi} \& {Ward}}{{Tanaka}
  et~al.}{2002}]{2002ApJ...565.1257T}
{Tanaka} H.,  {Takeuchi} T.,    {Ward} W.~R.,  2002, \apj, 565, 1257

\bibitem[\protect\citeauthoryear{{Ward}}{{Ward}}{1991}]{1991LPI....22.1463W}
{Ward} W.~R.,  1991, in Lunar and Planetary Institute Conference Abstracts
  {Horsehoe Orbit Drag}.
p.~1463

\bibitem[\protect\citeauthoryear{{Ward}}{{Ward}}{1997}]{1997Icar..126..261W}
{Ward} W.~R.,  1997, Icarus, 126, 261

\end{thebibliography}

\label{lastpage}
\end{document}